\documentclass[pre,twocolumn,showpacs]{revtex4-1}

\usepackage{graphicx}
\usepackage{amsmath}
\usepackage[dvips]{color}

\newcommand{\bnabla}{\mbox{\boldmath $\nabla$}}

\begin{document}

\title{A close look into the excluded volume effects within a double layer}

\author{Derek Frydel}
\affiliation{Institute of Physics, the Federal University of Rio Grande do Sul, 
PO Box 15051, 91501-970, Porto Alegre, RS, Brazil}
\author{Yan Levin}
\affiliation{Institute of Physics, the Federal University of Rio Grande do Sul, 
PO Box 15051, 91501-970, Porto Alegre, RS, Brazil}

\date{\today}

\begin{abstract}
We explore the effect of steric interaction on the ionic density distribution near a charged 
hard wall. For weakly charged walls, small particles, and monovalent ions the mean-field 
Poisson-Boltzmann equation provides an excellent description of the density profiles. For large 
ions and large surface charges, however, deviations appear.  To explore these, we use the density 
functional theory.  We find that local density functionals are not able to account for steric 
interactions near a wall. Based on the weighted density approximation we derive a simple 
analytical expression for the contact electrostatic potential which allows us to analytically 
calculate the differential capacitance of the double layer.  
 
\end{abstract}

\pacs{
}

\maketitle

\section{Introduction}

The standard Poisson-Boltzmann equation (PB) is accurate for dilute aqueous solutions containing 
small monovalent ions.  The first approximation of the PB theory is the mean-field account of the 
electrostatic interactions, where the discrete nature of ions is neglected and each ion interacts 
with the mean-potential generated by the local charge density distribution $\rho_c({\bf r})$ 
according to the Coulomb law, 
$\psi({\bf r})=(4\pi\epsilon)^{-1}\int d{\bf r}'\,\rho_c({\bf r}')/|{\bf r}'-{\bf r}|$, where 
$\epsilon$ is the dielectric constant.  This description clearly ignores ionic correlations.  The 
other approximation within the PB theory is the lack of internal structure of each ion:  an ion 
is a point in space characterized only by its charge.  Under this reduced description the ions
F$^-$, Cl$^-$, and NO$_3^-$ are indistinguishable, although their size~\cite{Le02} and 
polarizability~\cite{Frydel11} are obviously different.  The PB description, therefore, breaks 
down when the above reductions can no longer be justified.  In the present work we consider the 
weakly correlated limit so that the mean-field description of electrostatics holds, but we go beyond 
the standard PB equation by including the nonelectrostatic excluded volume effects~\cite{Le02}.  
In aqueous solutions the excluded volume interactions are further increased on account of 
hydration, the binding of water molecules to ions, so that the effective ionic radii are larger than the 
crystallographic ones~\cite{Ni59,LeSa09,Dzubiella10}.   Another effect of hydration, related 
directly to electrostatics, is the dielectric decrement \cite{Glueckauf64,David11,David12} 
associated with the decrease of the medium dielectric constant on account of polarization saturation
of a solvent constituting the hydration shell.  In the present work we do not 
consider this effect and focus strictly on the non-electrostatic excluded volume effects.   

The excluded volume effects reorganize the double layer so that a density profile is no longer 
monotonically decreasing, but it acquires oscillatory structure reflecting molecular composition 
of an electrolyte \cite{Marcia05}.  Clearly, the modifications of the density profile lead to 
modified electrostatic properties.  In the present study we take a close look into the excluded 
volume effects where ions are represented as hard spheres with the same diameter, 
the so-called, restricted primitive model.  Not only do we investigate how the ionic 
system responds to these effects under various conditions, but also how well various theories reproduce 
this behavior.  The hard-sphere interactions lead to the nonlocal effects, demanding 
nonlocal functionals to account for the oscillatory structure in the density profile.  The nonlocal 
treatments, such as the density functional theory (DFT), can be computationally demanding if the 
system is sufficiently complex.  Frequently in electrostatics, a variant of 
the local density approximation (LDA) is used, where the ideal gas entropy is substituted by 
that of the lattice-gas leading to the modified Poisson-Boltzmann equation (MPB) 
\cite{David97,Bazant09}.  With growing popularity of the MPB as a simple and tractable 
model, we feel that a careful and systematic study of this method is still lacking.  First of all,
the MPB is a local theory inappropriate for moderately and highly inhomogeneous fluids.  Furthermore, 
even as a theory of a homogeneous fluid it already fails at predicting the correct second 
virial term.   We consider another variant of the LDA based on the scaled particle theory
(SPT).  Surprisingly, although highly accurate for treating homogeneous fluids, it is less
accurate than the MPB, which is unreliable for homogeneous fluids. We compare all the results 
against the nonlocal fundamental measure density functional theory \cite{Rosenfeld89}.  
Our main focus is the region near and at the contact with the wall, where exact 
thermodynamic relations are known.  We investigate the scaling properties of the contact quantities 
and propose an analytical fit to the DFT data that permits an accurate prediction of these 
quantities, without resorting to numerical computations involving the entire region.

Throughout the paper we make frequent use of the initials:  DFT, LDA, MPB, and SPT.  
To avoid confusion we give a quick account of each.  DFT always refers to the fundamental
measure density functional theory as originally derived by Rosenfeld \cite{Rosenfeld89}.
This is the only nonlocal theory used in this work.  LDA refers to the general class of 
approximations where a density functional is defined locally based on a free energy of a 
homogeneous fluid.  MPB is the special example of the LDA, when the excluded volume effects 
are represented as that of a homogeneous lattice gas \cite{David97}.  And the SPT stands 
for the scaled particle theory and refers to the theoretical approach based on scaling a particle 
size \cite{Lebowitz59,Widom63,Andrews74,Corti}.  We use SPT to refer to another LDA 
approximation based on the SPT results for a uniform fluid.  In the paper two LDA 
approximations will be considered:  the MPB and the SPT.

The paper is organized as follows.  In Section II we discuss a reversible work of inserting a test 
particle into a hard sphere fluid, $W_{\rm hs}$, and emphasize its nonlocal character by 
deriving the exact result for inserting a point particle into a hard-sphere fluid.  We then 
introduce the fundamental density functional theory as an accurate nonlocal treatment of the
hard-sphere interactions.  In Section III we review the contact value theorem and explore its 
implications for the counterion distribution near a hard charged surface.  We point out that the LDA 
theories lead to contact relations different from the exact contact value theorem.  This is the
result of representing the overcrowding in the double-layer as the density "saturation".  
In Section IV we investigate the scaling of the quantity $\Delta_w$, which represents the
reversible work of brining the uncharged fluid particle from the bulk into contact with the wall. 
We propose the analytical formula for $\Delta_w$ that fits the DFT data points. 
In Section V we apply this formula to calculate the surface potential and the differential
capacitance.  Section VI finalizes this work with the concluding remarks.

\section{Preliminaries}

We write the free energy functional as having three separate contributions,
\begin{equation}
F = F_{\rm id} + F_{\rm c} + F_{\rm hs},
\end{equation}
the ideal gas contribution,
\begin{equation}
F_{\rm id} = k_BT\sum_{j=1}^K\int d{\bf r}\,\rho_j({\bf r})(\log\rho_j({\bf r})\Lambda^3-1),
\end{equation}
the mean-field electrostatic interactions,
\begin{equation}
F_{\rm c} = \frac{1}{8\pi\epsilon}
\int d{\bf r}\int d{\bf r}'\,\frac{[\rho_c({\bf r})+\rho_{\rm f}({\bf r})]
[\rho_c({\bf r}')+\rho_{\rm f}({\bf r}')]}{|{\bf r}'-{\bf r}|},
\end{equation}
and the excluded volume contributions, $F_{\rm hs}$, where all ions are taken to have the same 
diameter $\sigma$. $\rho_{\rm f}$ denotes a fixed charge of a macromolecule and in the present 
work we 
consider a charged wall, $\rho_f=\sigma_c\delta(x)$, where $\sigma_c$ is the surface charge.  
$\rho_c=\sum_{j=1}^Kq_j\rho_j$ is the charge density of mobile ions, where $q_j$ is the charge 
of an ion species $j$ and $K$ is the number of all ionic species.  Finally, $\Lambda$ is the 
de Broglie wavelength.  The equilibrium density minimizes the grand potential functional,
\begin{equation}
\Omega = F - \sum_{j=1}^K\mu_j\int d{\bf r}\,\rho_j({\bf r}),
\end{equation}
with respect to a density distribution, $\frac{\delta\Omega}{\delta \rho_j}=0$, and leads to
\begin{equation}
\rho_j({\bf r})=\rho_j^b e^{-\beta [q_j\psi({\bf r})+W_{\rm hs}({\bf r})-W_{\rm hs}^b]}.
\label{eq:rho_j}
\end{equation}
where $\psi=\frac{\delta F_{\rm c}}{\delta \rho_j}/q_j$ is the mean electrostatic potential,  
$W_{\rm hs}=\frac{\delta F_{\rm hs}}{\delta \rho}$, and $\rho_j^b$ is the bulk density of an 
ion species $j$.  

Physically, $W_{\rm hs}$ represents the reversible work of inserting a test particle at 
position ${\bf r}$ into a hard-sphere fluid \cite{Lebowitz59,Widom63,Andrews74,Corti}.  
If the system has a wall, the insertion performed far from the wall, 
$W_{\rm hs}({\infty})=W_{\rm hs}^b$ is identical with the excess (over the ideal gas) 
chemical potential.  
Alternatively, $W_{\rm hs}$ can be seen as a depletion interaction:  as a particle 
approaches a hard-wall, it feels itself being pushed towards the wall by the other particles. 
Application of the Poisson equation, $\epsilon\nabla^2\psi=-\rho_c$, leads to a kind of modified 
Poisson-Boltzmann equation,
\begin{equation}
\epsilon\nabla^2\psi = -\sum_{j=1}^K
\rho_j^b q_j e^{-\beta [q_j\psi({\bf r})+W_{\rm hs}({\bf r})-W_{\rm hs}^b]},
\end{equation}
where the surface charge $\sigma_c$ is determined by the boundary conditions at a charged wall,
$$
\epsilon\frac{\partial \psi}{\partial x} = -\sigma_c.  
$$
In order to obtain the potential, one still needs some kind of closure for $W_{\rm hs}$ that will 
determine the excluded volume effects.

\subsection{insertion of a point particle -- the exact case}

$W_{\rm hs}$ is a nonlocal quantity.  To see the origins of the non-locality we consider a 
simple case where the insertion work can be expressed as an exact density functional.   The case 
we refer to is the reversible work of inserting a {\it point test particle} into a hard-sphere fluid, 
$W_{\rm p}$.  The range of the hard-core interaction between the test point particle and the fluid 
hard-spheres is half the diameter, $\sigma/2$.  The reversible work is related 
to the insertion probability, $\Pi_{\rm p}=e^{-\beta W_{\rm p}}$.  For a given 
instantaneous configuration of a hard sphere fluid, the space is either occupied or is empty. 
The point particle can be inserted into the empty space between the hard spheres.  The 
instantaneous insertion probability, $\hat\Pi_{\rm p}$, is either zero, if we hit on an occupied 
space, or one, if a cavity is found, 
\begin{eqnarray}
\hat\Pi_{\rm p}({\bf r}) &=& 1 - \sum_{i=1}^N
\theta\big(\frac{\sigma}{2}-|{\bf r}_i-{\bf r}|\big)\nonumber\\
&=& 1 - \int d{\bf r}'\,\hat\rho({\bf r}')\theta\big(\frac{\sigma}{2}-|{\bf r}'-{\bf r}|\big),
\end{eqnarray}
where $\hat\rho({\bf r})=\sum_{i=1}^N\delta({\bf r}_i-{\bf r})$ is the density operator, $N$ is 
the total number of fluid particles, and $\theta(x)$ is the Heaviside step function.  Averaging over 
all the configurations, we obtain
\begin{equation}
\big\langle\hat\Pi_{\rm p}({\bf r})\big\rangle=\Pi_{\rm p}({\bf r}) = 1 - \int d{\bf r}'\,\rho({\bf r}')
\theta\big(\frac{\sigma}{2}-|{\bf r}'-{\bf r}|\big).
\end{equation}
The  reversible work of inserting a point particle at position ${\bf r}$ is
\begin{equation}
\beta W_{\rm p}({\bf r}) = -\log\bigg[1 - \int d{\bf r}'\,\rho({\bf r}')
\theta\big(\frac{\sigma}{2}-|{\bf r}'-{\bf r}|\big)\bigg].
\label{eq:W_sigma/2}
\end{equation}
The non-locality of  $W_{\rm p}$ is already evident in this simple example and shows 
that the work of inserting a point particle depends on the weighted average of the inhomogeneous 
density distribution.  The extent of the nonlocality is determined by the range of the hard-core 
interaction.  To obtain an expression for $W_{\rm hs}$, a reversible work associated with 
expanding a point particle to diameter $\sigma$ has to be considered, and 
$W_{\rm p}$ constitutes only a part of the entire quantity $W_{\rm hs}$.  For the homogeneous fluid, 
this procedure leads to the scaled particle theory (SPT) \cite{Lebowitz59}, in which the work of 
expansion to finite diameter $\sigma$ is interpolated between the two exact limits, 
$\lambda=\infty$ and $\lambda=\sigma/2$, where $\lambda$ is the range of the hard sphere 
interaction between a test and a fluid particle.  For an inhomogeneous fluid the derivation is 
more involved~\cite{Rosenfeld89}.

\subsection{density functional theory}

The ideas of the SPT applied to inhomogeneous fluids lead to the formulation of the fundamental 
measure density functional theory (DFT) \cite{Rosenfeld89}.  The DFT approximates the excess 
free energy due to the hard-sphere interactions through the free energy contribution, 
$\beta F_{\rm hs}[\rho]=\int d{\bf r}\,\Phi_{\rm hs}$, where 
\begin{equation}
\Phi_{\rm hs} = -\bar\rho_0\log(1-\bar\rho_3) + \frac{\bar\rho_1\bar\rho_2
-\bar{\boldsymbol\rho}_1\cdot\bar{\boldsymbol\rho}_2}{1-\bar\rho_3} 
+ \frac{\bar\rho_2}{24\pi}\frac{\bar\rho_2^2
-\bar{\boldsymbol\rho}_2\cdot\bar{\boldsymbol\rho}_2}{(1-\bar\rho_3)^2}.
\label{eq:Phi_ex}
\end{equation}
The hard sphere contributions are locally represented through weighted densities, instead of a 
physical density, as is the case for the LDA approximation, 
\begin{equation}
\bar\rho_{\alpha}(\bf r) = \int d{\bf r}'\,\rho({\bf r}')\omega_{\alpha}({\bf r}-{\bf r}')
\end{equation}
with $\omega_{\alpha}({\bf r}-{\bf r}')$ denoting a weight function \cite{Rosenfeld89,Roland10}.  
There are five weight functions which are obtained through decomposition of the Mayer-$f$ 
function into convolution products \cite{Korden12}.  
The two of the weight functions are vector quantities  and represent a distribution of normal 
to the sphere surface vectors.  In the uniform fluid they have no contribution.  The weight 
functions are given below,
\begin{equation}
\omega_2({\bf r}) = \delta(R-r),\,\,\,
\omega_3({\bf r}) = \theta(R-r),\,\,\,
{\boldsymbol\omega}_{2V}({\bf r}) = \frac{\bf r}{r}\delta(R-r),
\end{equation}
and $\omega_0({\bf r})=\omega_2({\bf r})/(4\pi R^2)$, 
$\omega_1({\bf r})=\omega_2({\bf r})/(4\pi R)$, and 
${\boldsymbol\omega}_{1V}({\bf r})={\boldsymbol\omega}_{2V}({\bf r})/(4\pi R)$, where 
$R=\sigma/2$.  Since we assume 
the size of all ions to be the same, the weight functions are the same for all the ionic species.  
The insertion work $W_{\rm hs}$ is defined as the 
functional derivative of the excess part of free energy,
\begin{equation}
\beta W_{\rm hs}({\bf r}) = \frac{\delta \beta F_{\rm hs}}{\delta\rho}=
\sum_{\alpha}\int d{\bf r}'\,
\frac{\partial\Phi_{\rm hs}({\bf r}')}{\partial\bar\rho_{\alpha}}\omega_{\alpha}({\bf r}'-{\bf r}).  
\label{eq:W_DFT}
\end{equation}
The DFT theory presented above accurately captures the dimensional 
crossover when the 3d fluid is confined to a quasi-2d space \cite{Rosenfeld96} --- if a degree of 
freedom is completely eliminated in the transverse direction, a 2d fluid is accurately recovered.

\begin{figure}[tbh]
\vspace{0.6cm}
\centerline{\resizebox{0.45\textwidth}{!}
{\includegraphics{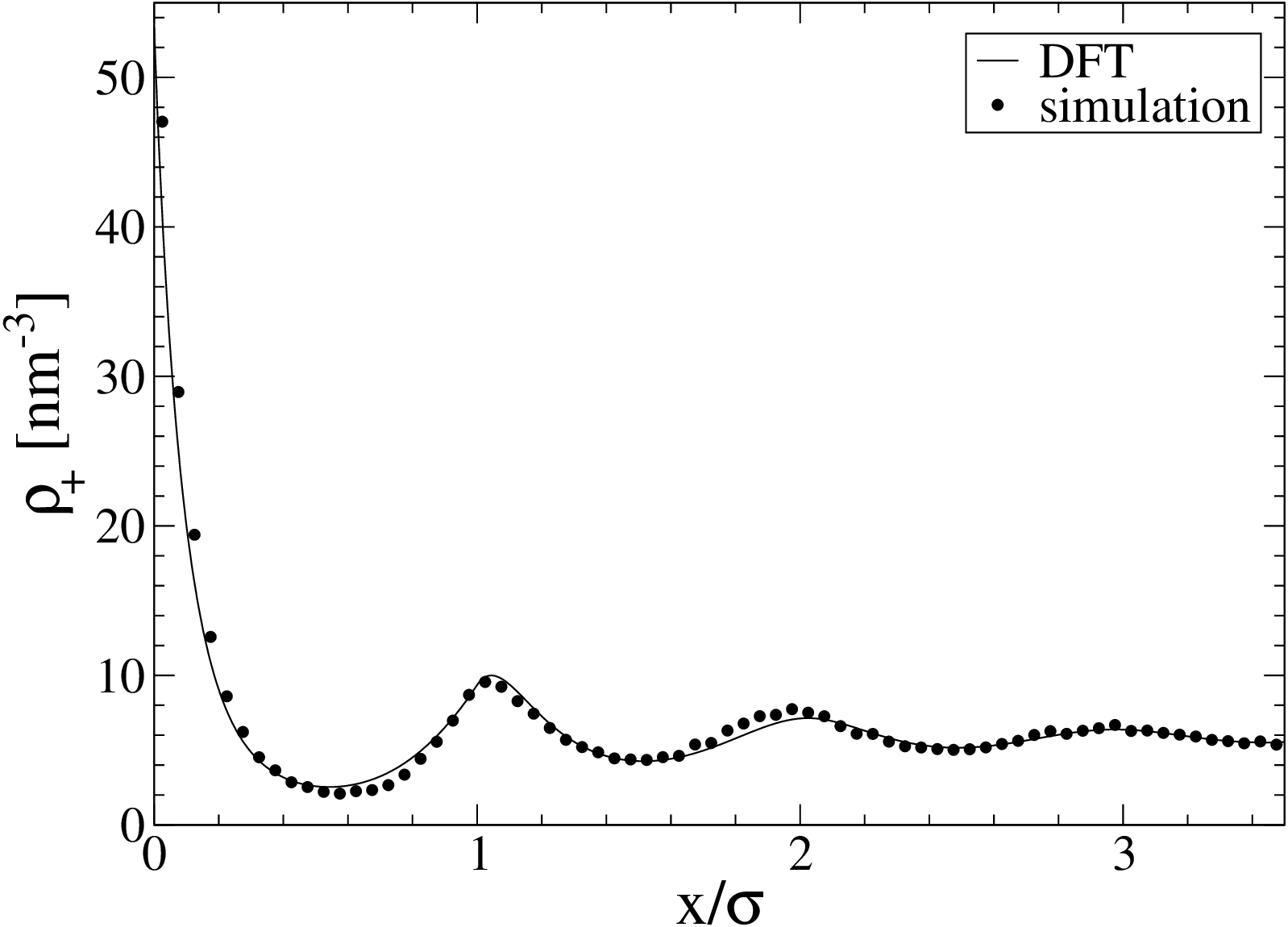}}}
\caption{The distribution of counterions near a wall.  The points are the results
of the Monte Carlo simulation of $1:1$ electrolyte between two parallel charged walls 
with the surface charge $\sigma_c=0.1{\rm C/m^2}$ and the separation $L=3.52{\rm nm}$. The 
remaining parameters are:   the particle diameter $\sigma=0.4{\rm nm}$, the salt concentration 
$c_s=9.5{\rm M}$, and the Bjerrum length $\lambda_B=e^2/\epsilon \sigma k_B T = 1{\rm nm}$, 
where $e$ is the elementary charge.  The solid line is the DFT prediction.  For technical 
details regarding the simulation see the reference \cite{Yan12}.}
\label{fig:rho_simulation}
\end{figure}

\subsection{local density approximation (LDA)}

Although highly accurate, see Fig. (\ref{fig:rho_simulation}), the numerical implementation of 
the fundamental measure density functional theory is quite involved.  To simplify the calculations, 
local density approximations (LDA) have often been invoked.  Within the LDA, the thermodynamic 
relations derived for a homogeneous fluid are applied {\sl locally} to an inhomogeneous situation.  
For example, within the LDA the insertion probability $W_{\rm hs}$  is supposed to be a function 
of the local density $\rho({\bf r})$ and is identified with the "local" excess chemical potential, 
$W_{\rm hs}({\bf r}) = W_{\rm hs}(\rho({\bf r}))$.  Clearly the specific form of the LDA will 
depend on the approximate form of $W_{\rm hs}$.  The MPB 
\cite{David97} relies on $W_{\rm hs}$ derived from the lattice-gas description of electrolyte, 
$\beta W_{\rm hs}=-\log(1-\eta)$, where $\eta=\pi\sigma^3\rho/6$ is the packing fraction.  
The SPT expression is 
$\beta W_{\rm hs}=-\log(1-\eta) + \frac{14\eta - 13\eta^2 + 5\eta^3}{2(1-\eta)^3}$.

\section{the contact value theorem}

In this work we focus on the region at a contact with a charged wall.  In this region the 
excluded volume effects are most pronounced since the ion concentration is highest, and 
for many problems the contact quantities are of primary interest.  This is the case for 
calculation of the differential 
capacitance \cite{Kornyshev07}, or the estimation of the strength of the electrostatic 
interactions between colloids trapped at an interface \cite{Frydel99}. The theoretical 
investigation in this region is facilitated by the existence of exact thermodynamic relations.  This 
section reviews some analytical results valid at a contact with a wall.  Throughout the paper we make 
frequent use of either the subscript or the superscript $w$ to indicate that a quantity is taken 
at contact with the wall.  In 
the same way we use $b$ to indicates bulk values.  We remind that the charge and number 
density are $\rho_c=\sum_{j=1}^Kq_j\rho_j$ and $\rho=\sum_{j=1}^K\rho_j$, respectively.    
Consequently, the bulk and contact value of the number density is 
$\rho_b=\sum_{j=1}^K\rho_j^b$ and $\rho_w=\sum_{j=1}^K\rho_j^w$, respectively.

\subsection{exact results}
The simplest expression of the contact value theorem relates the bulk pressure $P$ to the 
average momentum transferred at the hard-wall: $k_BT\rho_w=P$, where $\rho_w$ is the fluid 
density at the contact with the wall.  If the hard-wall has a surface charge $\sigma_c$ and the 
particles are hard-spheres with a central charge, then the contact value theorem takes the form 
\cite{Lebowitz79}, 
\begin{equation}
k_BT\rho_w = P +  \frac{\sigma_c^2}{2\epsilon}.
\label{eq:CVT}
\end{equation}
Within the DFT formalism $P_{\rm }=k_BT\rho_b(1+\eta_b+\eta_b^2)/(1-\eta_b)^3$ the pressure is given 
according to the 
scaled particle theory.  This result is a highly accurate approximation and the contact value theorem 
according to the DFT is essentially exact.  
The contact value relation is the result of the mechanical force balance condition,
\begin{equation}
-\rho\bnabla \big[k_BT\log\rho\big] = \rho_c\bnabla\psi + \rho\bnabla W_{\rm hs},
\label{eq:force_bal}
\end{equation}
where the thermodynamic force on the left-hand side is counterbalanced by the electrostatic force 
and the depletion force due to the collisions with other fluid particles on the right-hand side.

\begin{figure}[tbh]
\vspace{0.6cm}
\centerline{\resizebox{0.45\textwidth}{!}
{\includegraphics{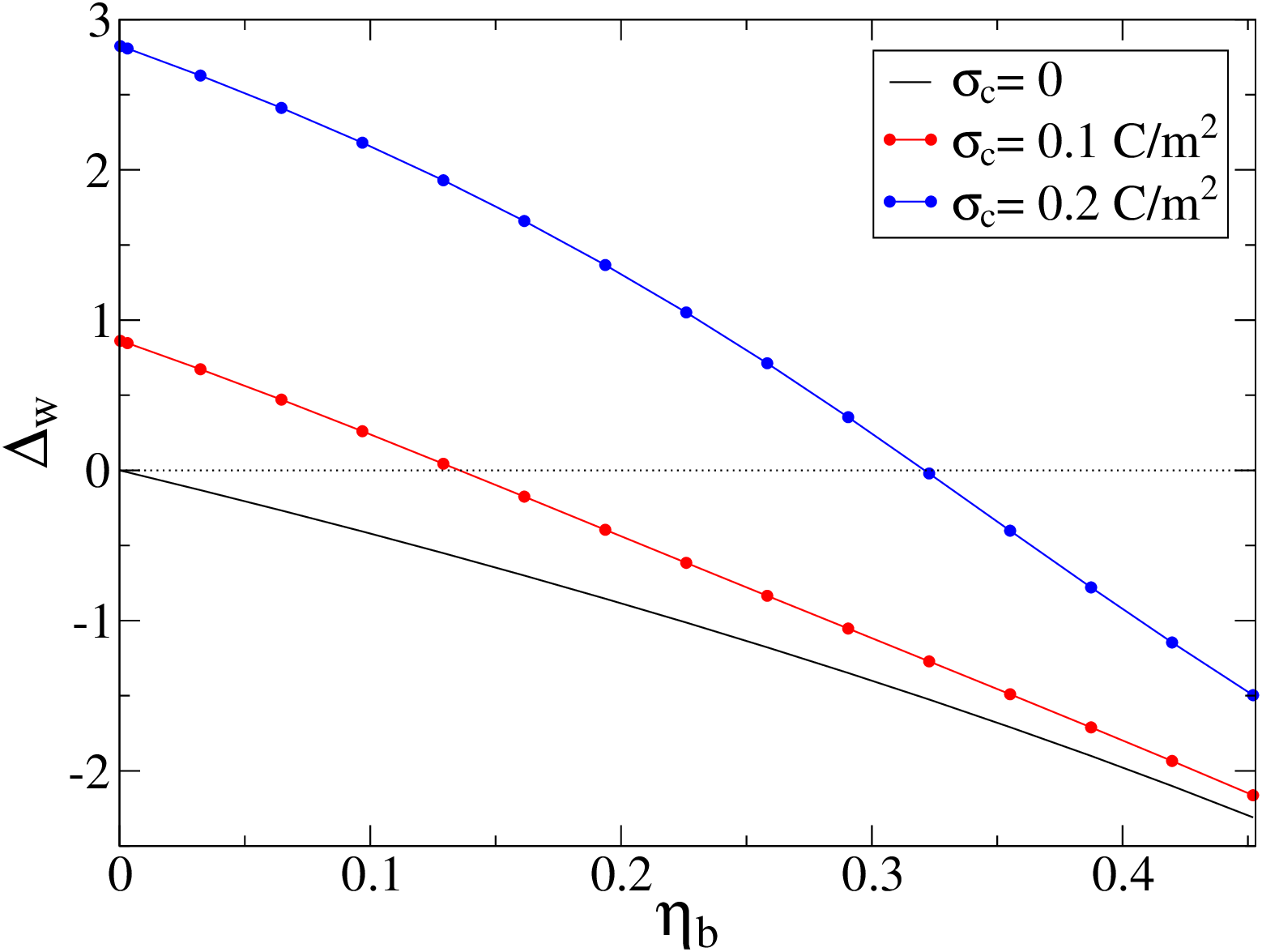}}}
\caption{The reversible work of dragging an uncharged test particle from bulk to the wall 
plotted as a function of the bulk packing fraction.  The data points are from the numerical DFT 
calculations for a symmetric salt $1:1$.    The bulk ion concentration is $\rho_b=2c_s$, the bulk packing fraction 
$\eta_b=\pi\sigma^3 c_s/3$, where $c_s$ is the bulk salt concentration.
The parameters are:  the particle diameter is $\sigma=0.8 {\rm nm}$, the Bjerrum length
is $\lambda_B=0.8 {\rm nm}$, and the bulk salt concentration is $c_s=0.1{\rm M}$.  Negative $\Delta_w$ 
signifies a favorable gain in free energy when bringing a hard sphere from the bulk to the wall.  }
\label{fig:Delta_vs_eta}
\end{figure}

The explicit expression for the contact density, using Eq.~(\ref{eq:rho_j}), is
\begin{equation}
\rho_w = \bigg[\sum_{j=1}^K\rho_j^b e^{-\beta q_j\psi_w}\bigg]e^{-\Delta_w}
\label{eq:rho_w}
\end{equation}
where we have introduced the quantity 
\begin{equation}
\Delta_w=\beta(W_{\rm hs}^w-W_{\rm hs}^b)= 
-\beta\int_0^{\infty}dx\,\frac{\partial W_{\rm hs}}{\partial x},
\label{eq:Delta_w}
\end{equation} 
representing the reversible work 
expanded to bringing a hard-sphere particle (with no electric charge) from the bulk electrolyte to 
the wall.  This is not to say that $\Delta_w$ is independent of electrostatic properties.  The work 
of bringing an uncharged hard-sphere to the wall depends on the ionic distribution within the 
double-layer that depends on the electrostatic interactions.  Substituting the density in 
Eq.~(\ref{eq:rho_w}) into the contact value theorem gives
\begin{equation}
k_BT\bigg[\sum_{j=1}^K\rho_j^b e^{-\beta q_j\psi_w}\bigg]e^{-\Delta_w} = 
P +  \frac{\sigma_c^2}{2\epsilon},
\label{eq:CVT11}
\end{equation}
For the case $\Delta_w=0$, $\psi_w$ is known exactly and is determined by the bulk quantities
and the surface charge.  The standard PB equation describes this scenario.  
A finite $\Delta_w$ modifies the contact potential.  $\Delta_w>0$ signifies that it is easier to 
insert a particle far away from the wall than at the contact.  This is intuitive because near the 
wall counterions are in excess.  Consequently the contact potential is larger than that obtained 
from the standard PB equation  ---  steric interactions deplete some ions from 
the first layer making it more difficult to neutralize the wall charge.  On the other hand, 
$\Delta_w<0$ signifies that it is easier to insert a particle at the wall than far from it.  
Although this seems counterintuitive, it can occur at low values of the surface charge 
$\sigma_c$ and/or large $\eta_b$.  To see this we consider the limit 
$\sigma_c\to 0$,
\begin{equation}
\lim_{\sigma_c\to 0}\Delta_w = -\log\bigg(\frac{P}{k_BT\rho_b}\bigg).
\label{eq:Delta_dilute}
\end{equation}
In this limit, $\Delta_w$ is always negative (since $P\geq k_BT\rho_b$), and the larger the bulk 
packing fraction $\eta_b$, the larger will be the modulus of $\Delta_w$.  At large ionic concentrations 
-- such as, for example, found in ionic liquids -- the range of $\sigma_c$ where $\Delta_w$ is 
negative can be significant.  In Fig.~(\ref{fig:Delta_vs_eta}) we plot $\Delta_w$ as a 
function of $\eta_b$ for different values of the surface charge.  The crossover where $\Delta_w$ 
changes sign to negative is pushed to larger $\eta_b$ as $\sigma_c$ becomes larger.  Eventually 
the crossover is suppressed as $\eta_b$ passes the freezing transition 
(at $\eta_b\approx 0.49$).  

\subsection{local density approximation (LDA)}

Within the local density approximation the insertion work is a local quantity, 
\begin{equation}
W_{\rm hs} = \frac{\partial \Phi_{\rm hs}}{\partial\rho}\Big|_{\rho=\rho({\bf r})},
\end{equation}
and the force density due to the hard-sphere interactions can be related to the gradient of
a local pressure, 
\begin{equation}
-\rho\bnabla W_{\rm hs} = -\bnabla\Bigg[\rho\frac{\partial\Phi_{\rm hs}}{\partial\rho}
-\Phi_{\rm hs}\Bigg]=-\bnabla [P-k_BT\rho_b].
\end{equation}
The force balance equation in Eq.~(\ref{eq:force_bal}) becomes
\begin{equation}
\rho_c\bnabla\psi = -\bnabla P,
\end{equation}
leading, after integrating over a half-space, to the contact relation,
\begin{equation}
P_w = P +  \frac{\sigma_c^2}{2\epsilon},
\label{eq:CVT_lda}
\end{equation} 
where $P_w$ is the local pressure at contact with the wall.  The notion of a local pressure can be 
somewhat misleading, since the pressure is normally associated with bulk properties.  
It is only sensible to talk about a local pressure within the framework of a local density 
approximation.  

\begin{figure}[tbh]
\vspace{0.6cm}
\centerline{\resizebox{0.45\textwidth}{!}
{\includegraphics{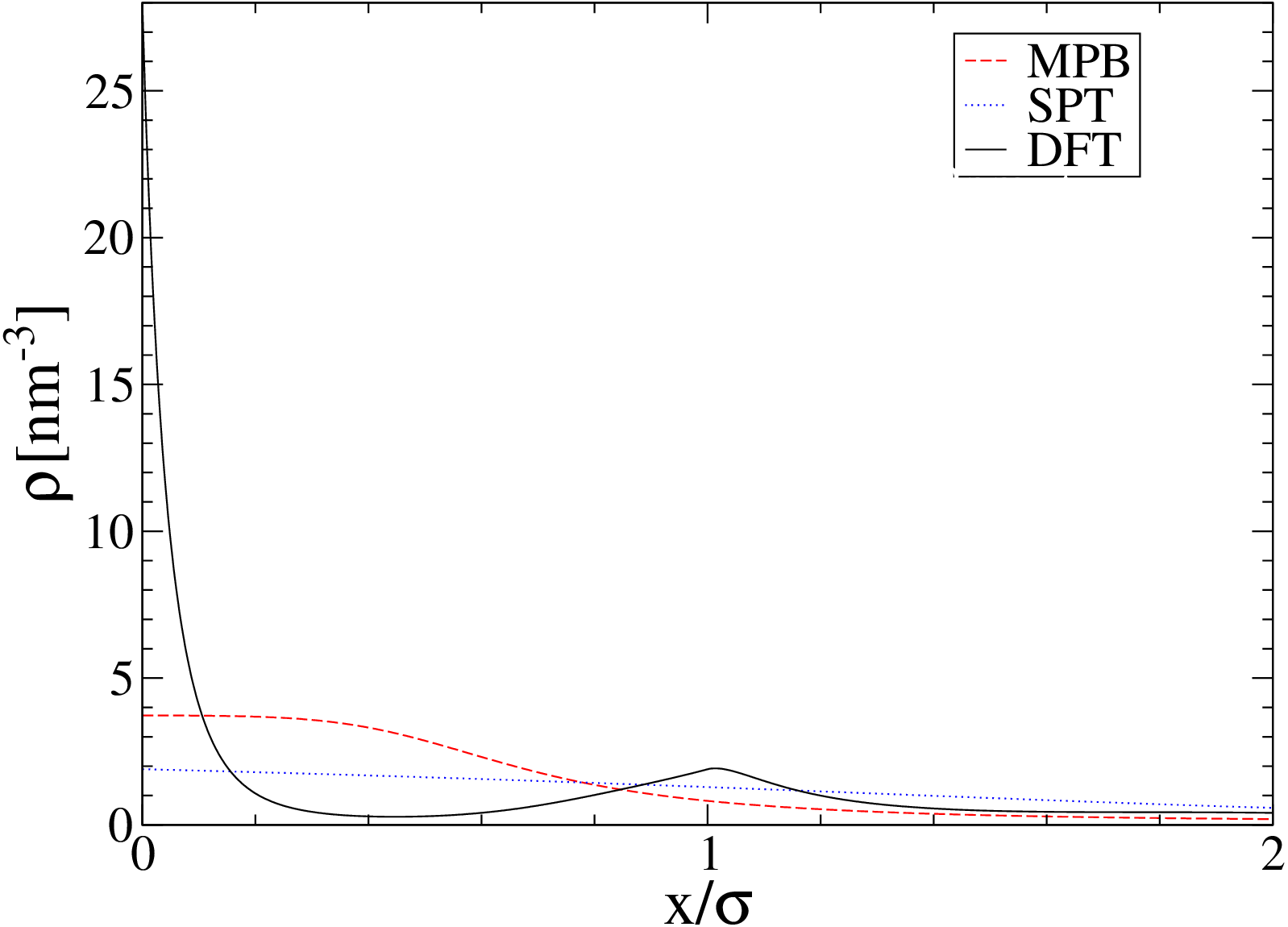}}}
\caption{The density profiles near a hard charged wall obtained using the DFT and the two LDA
methods: the MPB and the SPT.  
The parameters are: $c_s=0.1{\rm M}$, $\sigma_c=0.4{\rm C/m^2}$, 
$\lambda_B=0.72{\rm nm}$, and $\sigma=0.8{\rm nm}$.  The system represents a symmetric 
$1:1$ salt.}
\label{fig:rho_DFT}
\end{figure}

Going back to the exact contact value theorem in Eq.~(\ref{eq:CVT}) shows that the correct 
thermodynamic result is obtained only if $P_{\rm id}=k_BT\rho_b$.  
Paradoxically, if one tries to improve on this result and go beyond the PB description by including 
an excess free energy due to electrostatic or steric correlations using a LDA formalism, one 
obtains less accurate contact value density than if the correlations are neglected altogether.  The 
unphysical contact value relation for the LDA treatment is a consequence of an unphysical density 
profile.  The LDA approximation fails to capture the oscillating discrete structure of a fluid, see 
Fig.~(\ref{fig:rho_DFT}).  
As seen in that figure, the LDA approximations account for overcrowding through "density 
saturation", while a realistic description shows overcrowding as layering of a fluid near a wall.  
It is tempting to assume that the "saturation" observed in a density for the LDA models represents 
a weighted rather than a physical density profile,
\begin{equation}
\rho^{\rm LDA} \rightarrow \int_0^{\infty} dx'\rho(x')\, \omega(x'-x),
\end{equation}
where $\omega(x'-x)$ is some weight function, so that oscillations are completely smoothed out. 
Such conjecture, however, entails that $\psi$ 
obtained from the LDA is also not physical but only a weighted quantity.  We, therefore, consider that 
all quantities produced by the LDA are what they are, and that the saturation of a density profile 
is an artifact of the LDA treatment.  To test accuracy of the LDA one should rather consider 
other quantities such as $\psi$, which is of direct interest to most problems of electrostatics.

Below we look into details of two LDA models.  As mentioned earlier, the SPT results are highly 
accurate for homogeneous fluids, producing the exact second virial coefficient. Furthermore, the 
third and higher order virial coefficients are very accurate,
\begin{equation}
\beta P_{\rm spt} = \frac{\rho(1+\eta+\eta^2)}{(1-\eta)^3} 
= \rho\Big[1 + 4\eta + \dots\Big].
\end{equation}
On the other hand, the lattice-gas model (from which the MPB is derived) already fails at the 
second virial coefficient,
\begin{equation}
\beta P_{\rm lg}=-\frac{6}{\pi\sigma^3}\log(1-\eta) = \rho\Big[1 + \frac{1}{2}\eta + \dots\Big],
\end{equation}
and is unfit for treating homogeneous fluids.

\begin{figure}[tbh]
\vspace{0.6cm}
\centerline{\resizebox{0.45\textwidth}{!}
{\includegraphics{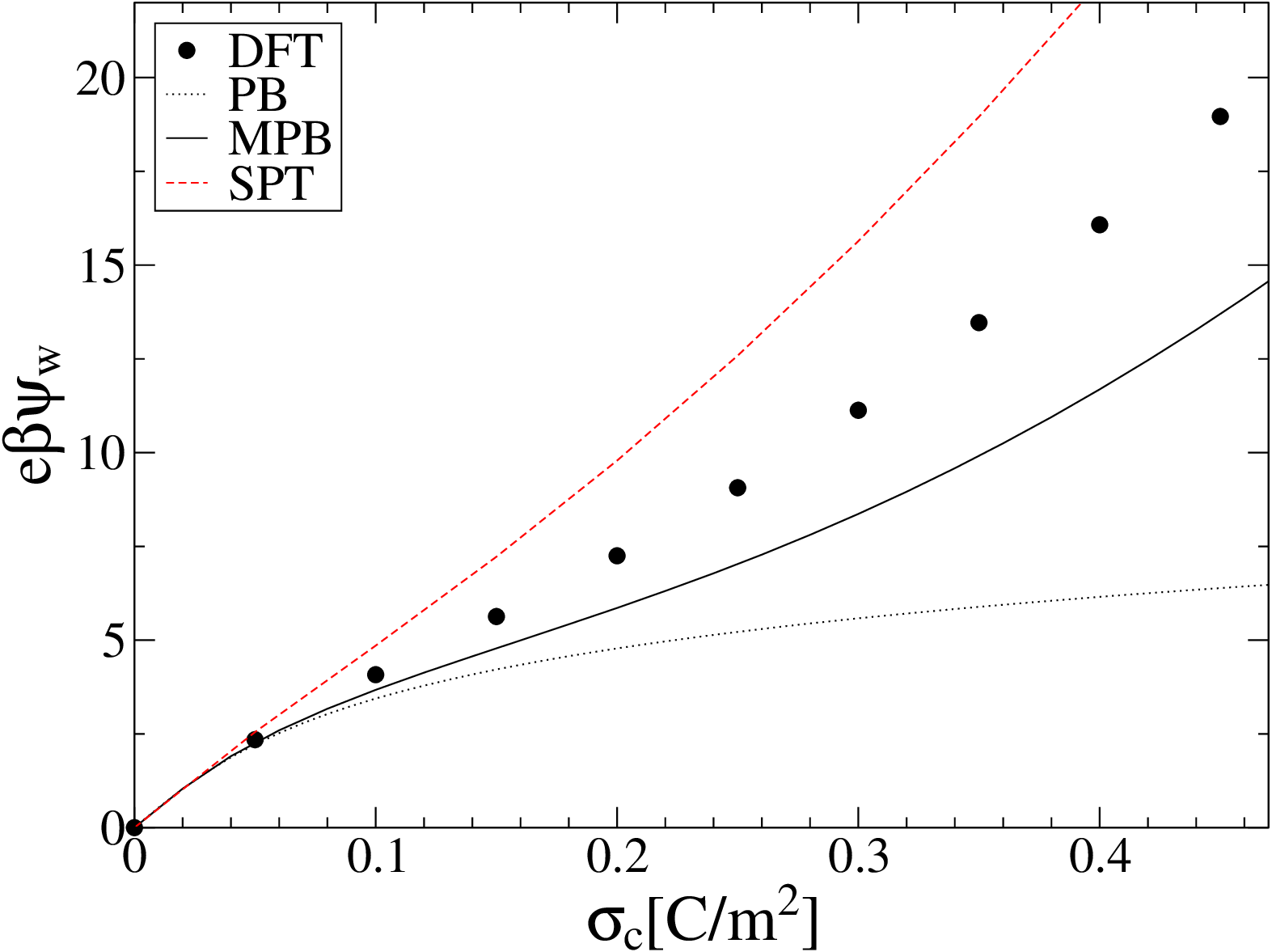}}}
\caption{The contact potential as a function of the surface charge.   The parameters are: 
$c_s=0.1{\rm M}$, $\lambda_B=0.72{\rm nm}$, and $\sigma=0.8{\rm nm}$.  
The system represents a symmetric $1:1$ salt. }
\label{fig:phi_vs_sigma_c_LDA}
\end{figure}

On the other hand, when the LDA is applied to inhomogeneous fluids, it is hard to predict what 
will happen.   In Fig.~(\ref{fig:phi_vs_sigma_c_LDA}) we plot the contact potential for various
models.  First we note that the difference between the PB and the DFT results.  The finite size 
effects lead to an increase of the surface potential, and the larger the surface charge, the larger
the discrepancy, since more counterions are crowded into the double-layer.  Next we discuss the
LDA models.  The MPB model underestimates while the LDA based on the SPT 
overestimates the DFT result.  

\begin{figure}[tbh]
\vspace{0.6cm}
\centerline{\resizebox{0.45\textwidth}{!}
{\includegraphics{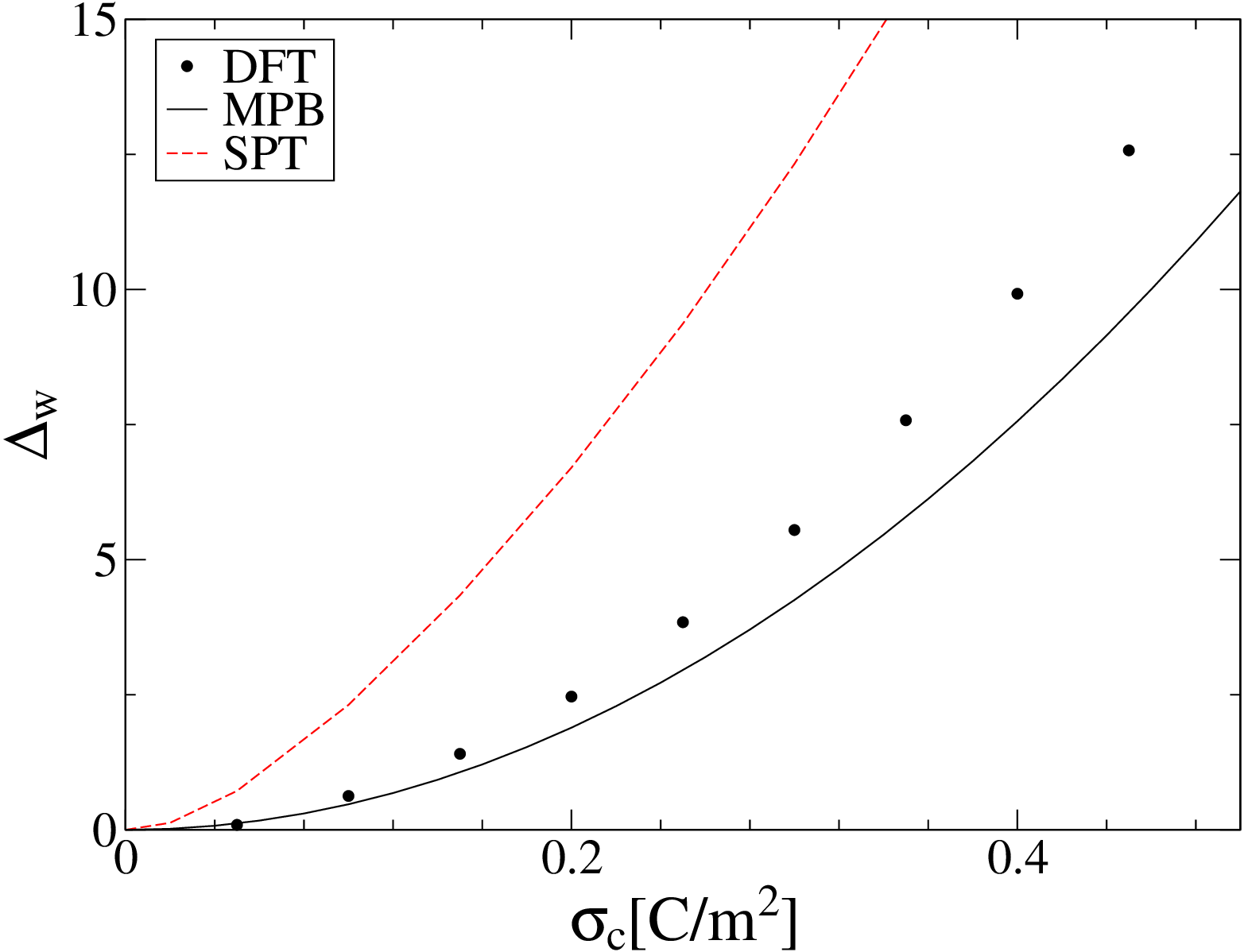}}}
\caption{The reversible work of bringing an uncharged test particle from the bulk to the wall 
as a function of the surface charge.   The parameters are:   
$c_s=0.1{\rm M}$, $\lambda_B=0.72{\rm nm}$, and $\sigma=0.8{\rm nm}$.  
The system represents a symmetric $1:1$ salt. }
\label{fig:Delta_LDA3}
\end{figure}

We next consider the quantity $\Delta_w$ as defined in Eq.~(\ref{eq:Delta_w}).  
As shown in Fig.~(\ref{fig:Delta_LDA3}) the MPB reproduces this quantity better than 
the STP.  The advantage of the MPB model is its analytical tractability, and according to this 
model $\Delta_w$ yields a very simple expression,
\begin{equation}
\Delta_w = \frac{\beta\pi\sigma_c^2\sigma^3}{12\epsilon},
\label{eq:Delta_mpb}
\end{equation}
with $\Delta_w$ parabolic in $\sigma_c$.  

\section{Scaling  of the quantity $\Delta_w$}

All the calculations in this work are carried out for the $1:1$ symmetric salt, so that the bulk 
number density is $\rho_b=2c_s$, where $c_s$ denotes the bulk salt concentration.  In this 
part of the study we investigate the scaling of the quantity $\Delta_w$.  $\Delta_w$  is a 
function of three dimensionless variables:  $\eta_b=\pi\sigma^3\rho_b/6$, 
$\sigma_c^*=\pi\sigma_c\sigma^2/(4e^2)$, and $\lambda_B^*=\lambda_B/\sigma$, corresponding 
to volume, surface, and length, respectively.  $\sigma_c^*$ represents the packing 
fraction of counterions projected on a 2d plane.  The 2d solid phase transition is at 
$\eta_{2d}>0.72$, and the maximum possible packing fraction is at $\eta_{2d}=0.90$.  If 
$\sigma_c^*<0.90$, potentially all the counterions can collapse onto the charged surface, 
if attraction is sufficiently strong (although $\sigma_c^*>0.72$ would involve a phase 
transition as well).   If, on the other hand, $\sigma_c^*>0.9$, collapse onto a single 
plane is prevented and a second fluid layer must form. 
In the reduced units, the contact value theorem is
\begin{equation}
\beta\nu P+\frac{16}{3}\lambda_B^*\sigma_c^{*2} = \eta_b\cosh\phi_w e^{-\Delta_w},
\label{eq:CVT11}
\end{equation}
where $\phi=e\beta\psi$ is the reduced potential, and $\nu=\pi\sigma^3/6$ is the volume
of a single fluid particle.

According to the LDA, $\Delta_w$ is a function of two independent parameters, $\eta_b$ and 
$\lambda_B^*\sigma_c^{*2}$.  Later, as we investigate $\Delta_w$ for various parameters 
using the DFT, we do not find this reduced description to be confirmed, and we conclude
that it is an artifact of the LDA.  Furthermore, this implies that the scaling of the contact 
potential $\psi_w$ according to the LDA description is described by two independent 
parameters.  The two parameter scaling of the LDA can be ascertained from 
the contact relation in Eq.~(\ref{eq:CVT_lda}).

\subsection{dilute limit}

\begin{figure}[tbh]
\vspace{0.6cm}
\centerline{\resizebox{0.45\textwidth}{!}
{\includegraphics{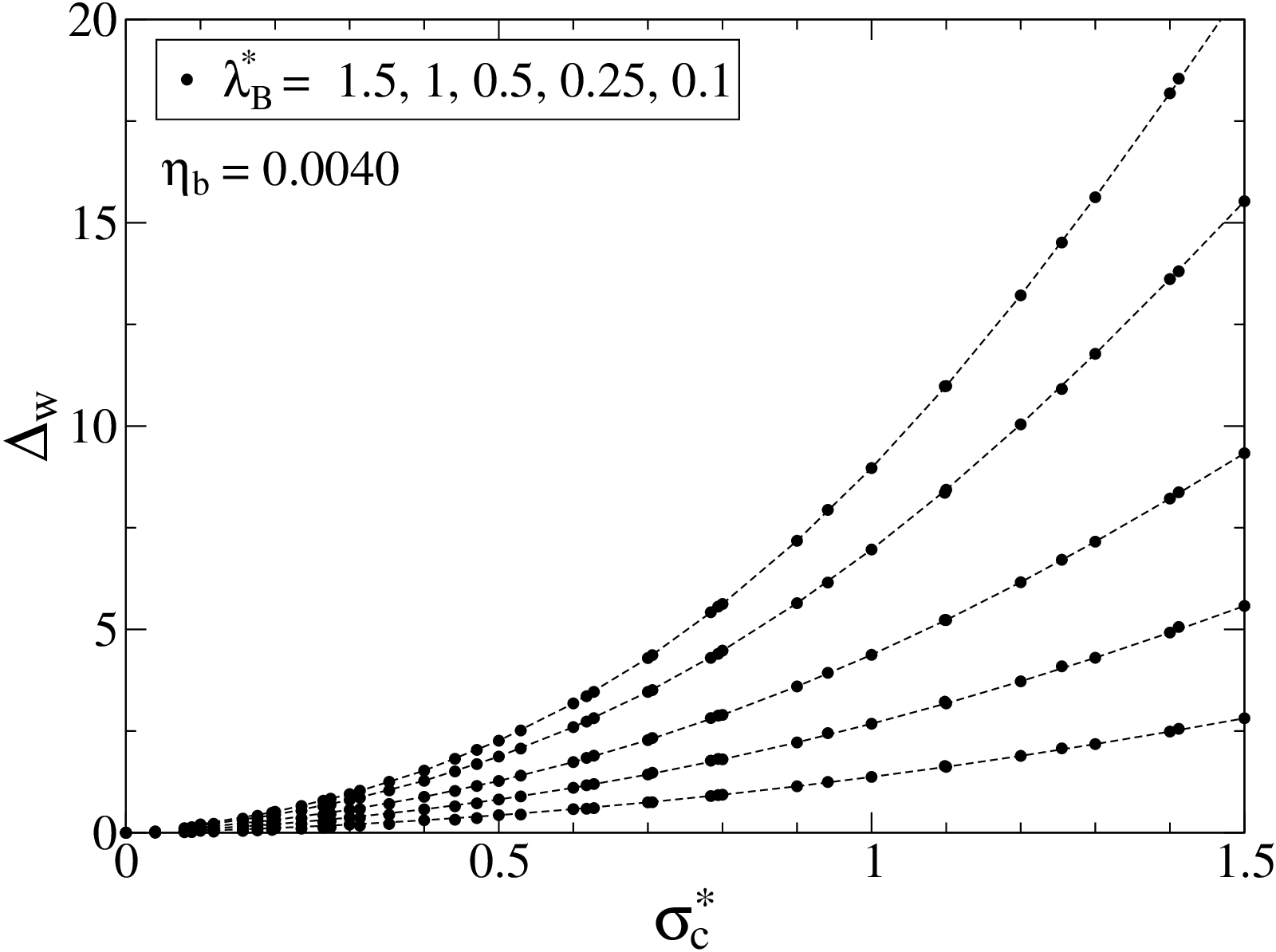}}}
\caption{$\Delta_w$ as a function of the surface charge. 
The packing fraction is fixed at $\eta_b=0.0040$ (the dilute limit).  
The points are the DFT results and the lines
are given by Eq. (\ref{eq:Delta0}). }
\label{fig:Delta_vs_sigma_c_A}
\end{figure}

\begin{figure}[tbh]
\vspace{0.6cm}
\centerline{\resizebox{0.45\textwidth}{!}
{\includegraphics{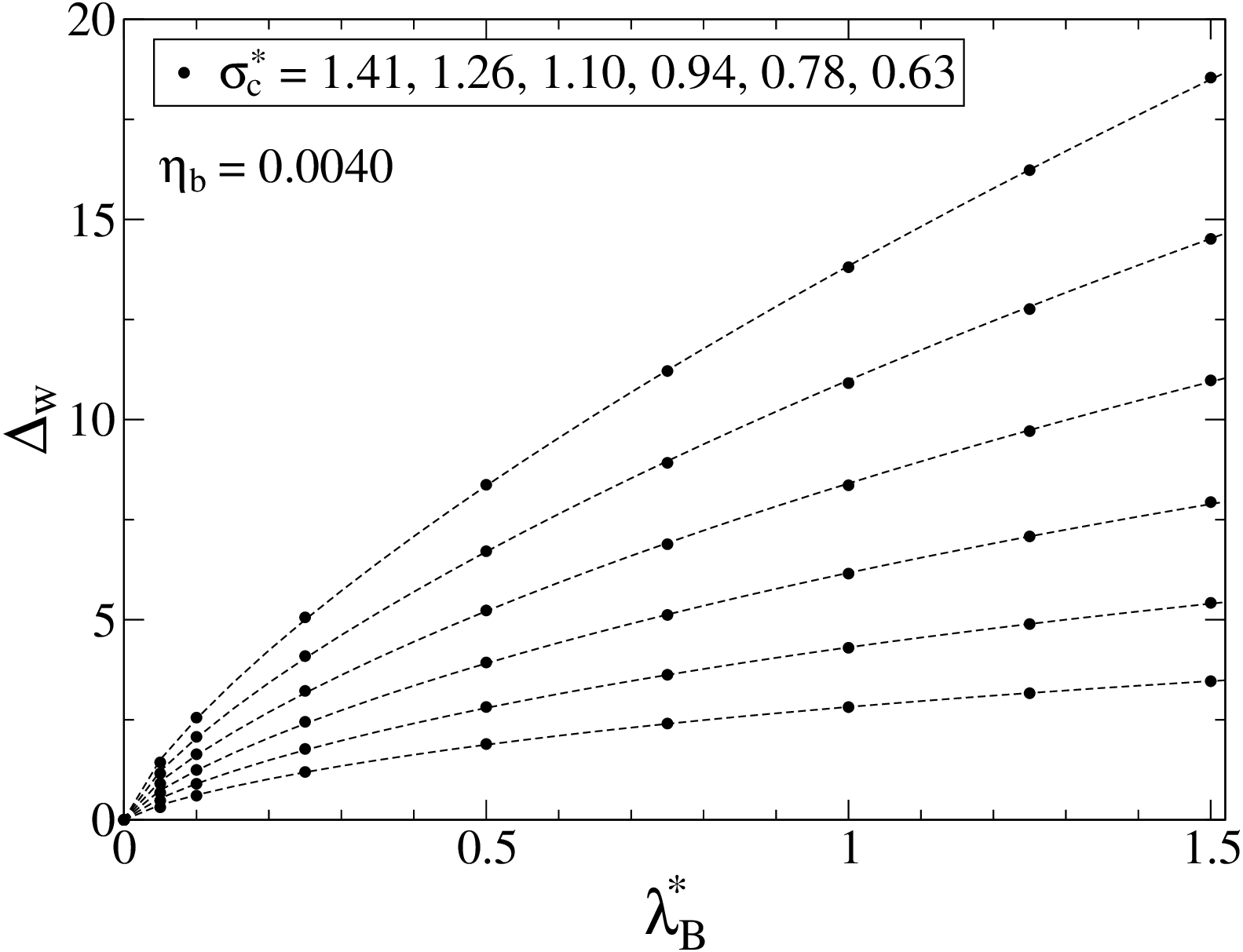}}}
\caption{$\Delta_w$ as a function of the Bjerrum length.  The 
packing fraction is fixed at $\eta_b=0.0040$ and represents (the dilute limit).  
The points are the DFT results and the lines
are given by Eq. (\ref{eq:Delta0}). }
\label{fig:Delta_vs_lambda}
\end{figure}

To simplify things we first consider the dilute limit, so as to suppress the 
contributions from the bulk eliminating the dependence on $\eta_b$,  
$\lim_{\eta_b\to 0}\Delta_w=\Delta_w^0\equiv\Delta_w^0(\sigma_c^*,\lambda_B^*)$.  
The main results are in Fig.~(\ref{fig:Delta_vs_sigma_c_A}) and 
Fig.~(\ref{fig:Delta_vs_lambda}).  The figures show the scaling behavior of $\Delta_w$.  The 
general observation is that the scaling with the surface charge is stronger than that with 
the Bjerrum length.  This can be understood since the surface charge controls both the interactions 
between the surface and counter charges and the counterion concentration.  On the 
other hand, the Bjerrum length regulates only the strength of the ionic interactions.   

Our aim is to propose a fit which reproduces the scaling of $\Delta_w$.  The desirable 
procedure would be to suggest an expression based on a physically motivated model.  
Unfortunately, an accurate physically motivated fit could only be found for a limited range.  To 
accurately parametrize the entire range we had to resort to a fitting procedure where the choice 
of a functional form was based on simplicity and convenience.   The following functional form 
was found convenient, 
$\tilde\Delta_w^0=a(\sigma^*)\{\lambda_B^*/[1+b(\sigma^*)\lambda_B^*]\}^\alpha$.   
Fully parametrized, this form reads
\begin{eqnarray}
\Delta_w^0 &=& (2.28\sigma_c^*+5.53\sigma_c^{*2})\nonumber\\
&\times&\Bigg[\frac{\lambda_B^*}{1+(1-1.21\sigma_c^*+0.375\sigma_c^{*2})\lambda_B^*}\Bigg]^{3/4}.
\label{eq:Delta0}
\end{eqnarray}
The accuracy of this fit is demonstrated in Figs. (\ref{fig:Delta_vs_sigma_c_A}) and 
Fig.~(\ref{fig:Delta_vs_lambda}) in comparison to the DFT data points.  
We also compare this formula with the simple expression from the MPB model in 
Eq.~(\ref{eq:Delta_mpb}) which in reduced units reads
\begin{equation}
\Delta_w = \frac{16}{3}\sigma_c^{*2}\lambda_B^*.
\label{eq:Delta_mpb2}
\end{equation} 
There is a curious similarity between the square term in our fit $=5.53$ and the coefficient $16/3$
in the MPB expression.  If this similarity is more
than coincidental, it would suggest that although reduced, the MPB description accurately captures 
some aspects of the hard-sphere interactions.

\subsection{finite concentrations}

\begin{figure}[tbh]
\vspace{0.6cm}
\centerline{\resizebox{0.45\textwidth}{!}
{\includegraphics{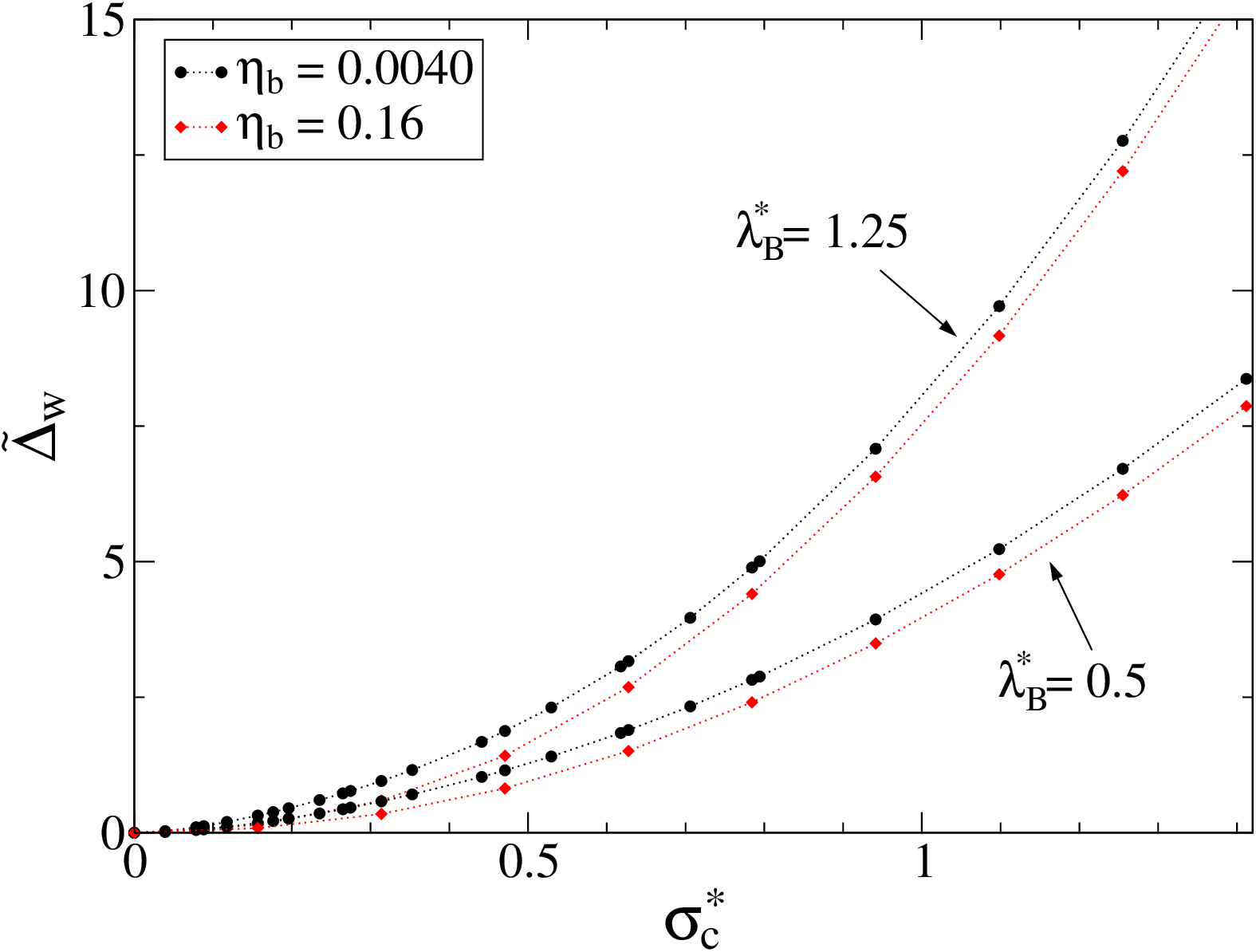}}}
\caption{$\tilde\Delta_w$ as a function of the surface charge for different Bjerrum lengths and 
packing fractions.  }
\label{fig:Delta_vs_sigma_c_B}
\end{figure}

We now turn to investigate the contributions of finite concentrations.  We suggest the 
following formula, 
\begin{equation}
\tilde\Delta_w = \Delta_w + \log\Bigg(\frac{P}{k_BT\rho_b}\Bigg) = \Delta_w^0 - \theta(\eta_b,\lambda_B^*),
\label{finitet}
\end{equation}  
so that $\Delta_w$ as a function of $\sigma_c^*$ is merely shifted by the quantity $\theta$.     
This fit is suggested by Fig.~(\ref{fig:Delta_vs_sigma_c_B}) where  
we compare $\Delta_w$ for finite $\eta_b$ with that for a dilute limit.  
In the relevant regime $\Delta_w$ as a function of 
$\sigma_c^*$ is offset by a value independent of $\sigma_c^*$.  We find the following
accurate fit to the function $\theta$,
\begin{equation}
\theta(\eta_b,\lambda_B^*) =  
1.75\eta_b + \eta_b^2\bigg[\frac{10.8\lambda_B^*-1}{0.21 +\lambda_B^*}\bigg].
\label{eq:}
\end{equation}
We subtract $-\log(P/k_BT\rho_b)$ from $\Delta_w$ so that the quantity $\tilde\Delta_w$ is always
positive, see Eq.~(\ref{eq:Delta_dilute}) and Fig.~(\ref{fig:Delta_vs_eta}).
To ensure that $\theta < \Delta^0_w$, so that $\tilde\Delta_w>0$ for all values of $\eta_b$, 
we apply the form,
\begin{equation}
\tilde\Delta_w = \Delta_w^0 e^{-\theta/\Delta_w^0}.
\label{eq:Delta_w_complete}
\end{equation}

\section{applications}

\begin{figure}[tbh]
\vspace{0.6cm}
\centerline{\resizebox{0.45\textwidth}{!}
{\includegraphics{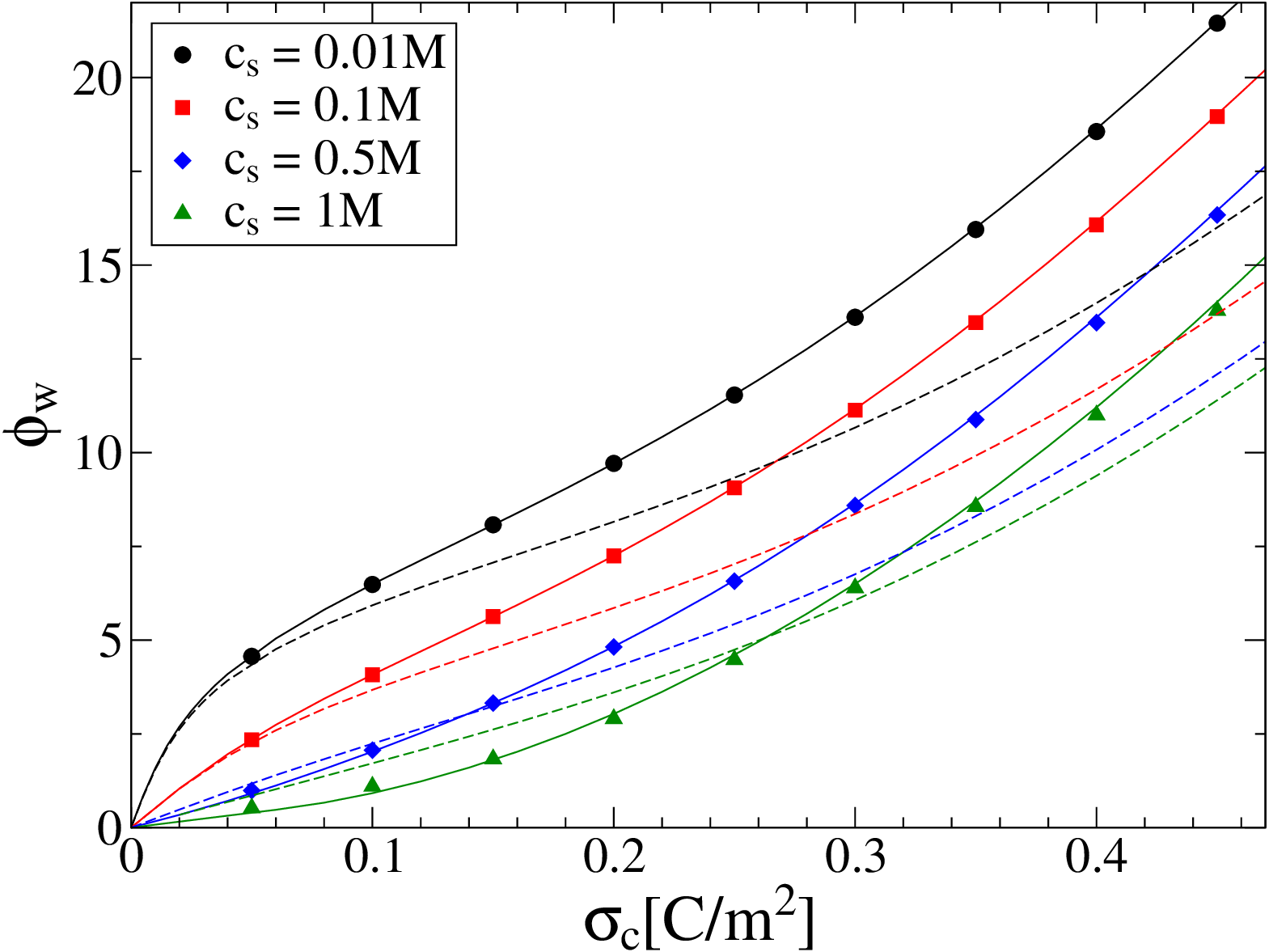}}}
\caption{Surface potential as a function of the surface charge.
The parameters are: $\lambda_B=0.72{\rm nm}$ and $\sigma=0.8{\rm nm}$.  The solid lines
correspond to the analytical expression based on Eq.~(\ref{eq:coshphi})
and the dashed lines to the MPB model.  }
\label{fig:phi_vs_sigma_c_A}
\end{figure}

\begin{figure}[tbh]
\vspace{0.6cm}
\centerline{\resizebox{0.45\textwidth}{!}
{\includegraphics{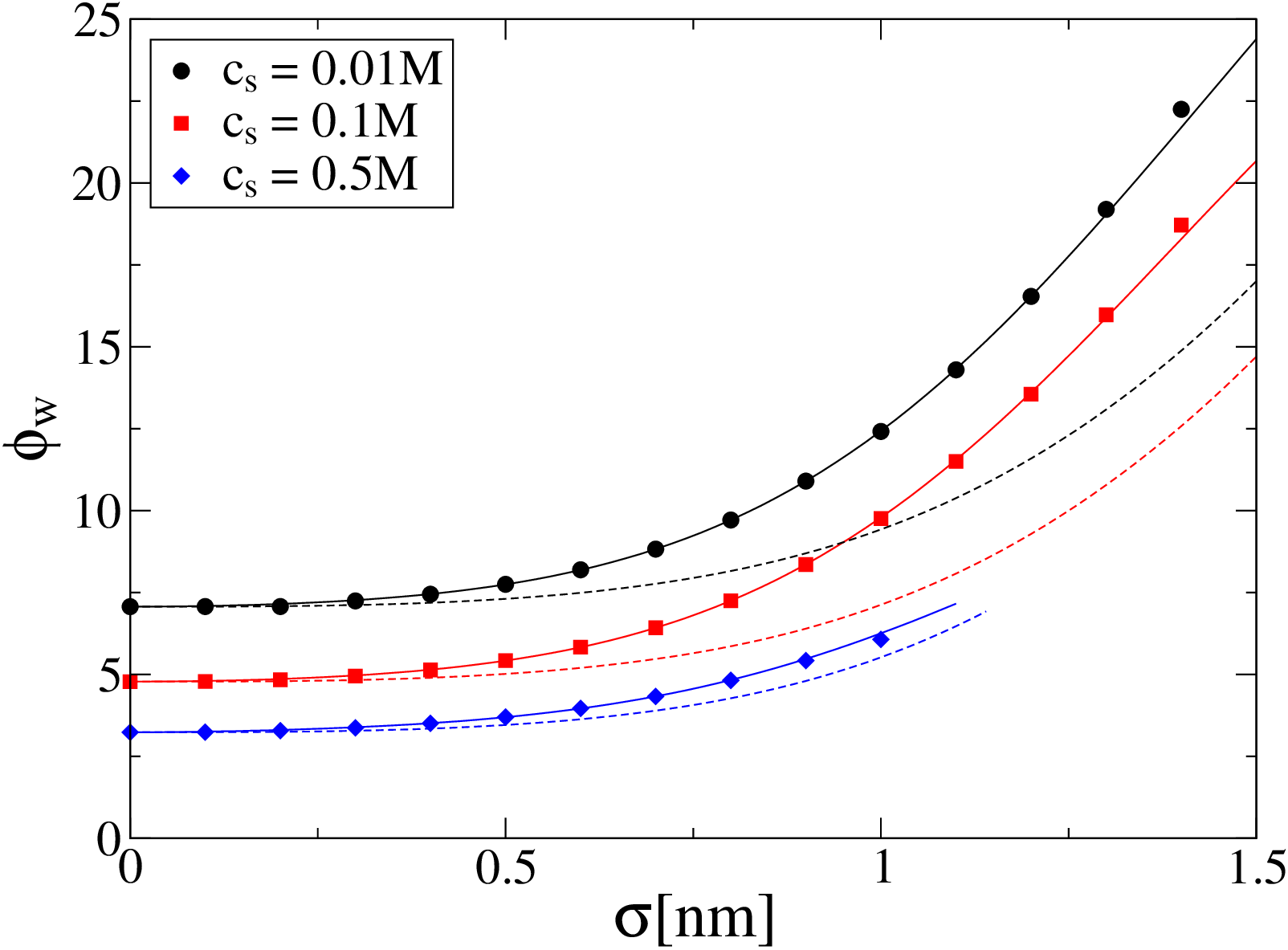}}}
\caption{Surface potential as a function of the particle diameter.
The parameters are: $\sigma_c=0.2{\rm C/m^2}$ and $\lambda_B=0.72{\rm nm}$.  The 
solid lines correspond to our analytical expression and the dashed lines to the MPB model.  Since 
the hard sphere fluid freezes at $\eta_b \simeq 0.49$, we interrupt the curve for $c_s=0.5M$. }
\label{fig:phi_vs_sigma_c_B}
\end{figure}

We now have the complete expression for $\Delta_w$,
\begin{equation}
\Delta_w = -\log\bigg(\frac{\beta P}{\rho_b}\bigg) + \tilde\Delta_w,
\end{equation}
where $\tilde\Delta_w$ is given in Eq.~(\ref{eq:Delta_w_complete}).
To obtain the contact potential, we insert $\Delta_w$ into   
Eq. (\ref{eq:CVT11}) to get
\begin{equation}
\cosh\phi_w = 
e^{\tilde\Delta_w}\Bigg(1+\frac{16}{3}\frac{\lambda_B^*\sigma_c^{*2}}{\nu\beta P}\Bigg).
\label{eq:coshphi}
\end{equation}
In Fig.~(\ref{fig:phi_vs_sigma_c_A}) and Fig.~(\ref{fig:phi_vs_sigma_c_B}) we 
compare the exact DFT results for the contact potential with the ones obtained using the 
analytical scaling function Eq.~(\ref{eq:coshphi}) and with the predictions of the MPB model.  
The scaling form derived in the present paper agrees very well with the contact 
potential obtained using the exact DFT, even at densities all the way to the solid
phase transition.  For largest particle sizes, Fig.~(\ref{fig:phi_vs_sigma_c_B}) shows
that the scaling function slightly disagrees with the DFT results.  This happens when 
$\sigma_c^*>2$ and a third layer of condensed counterions starts to form.  

\begin{figure}[tbh]
\vspace{0.6cm}
\centerline{\resizebox{0.45\textwidth}{!}
{\includegraphics{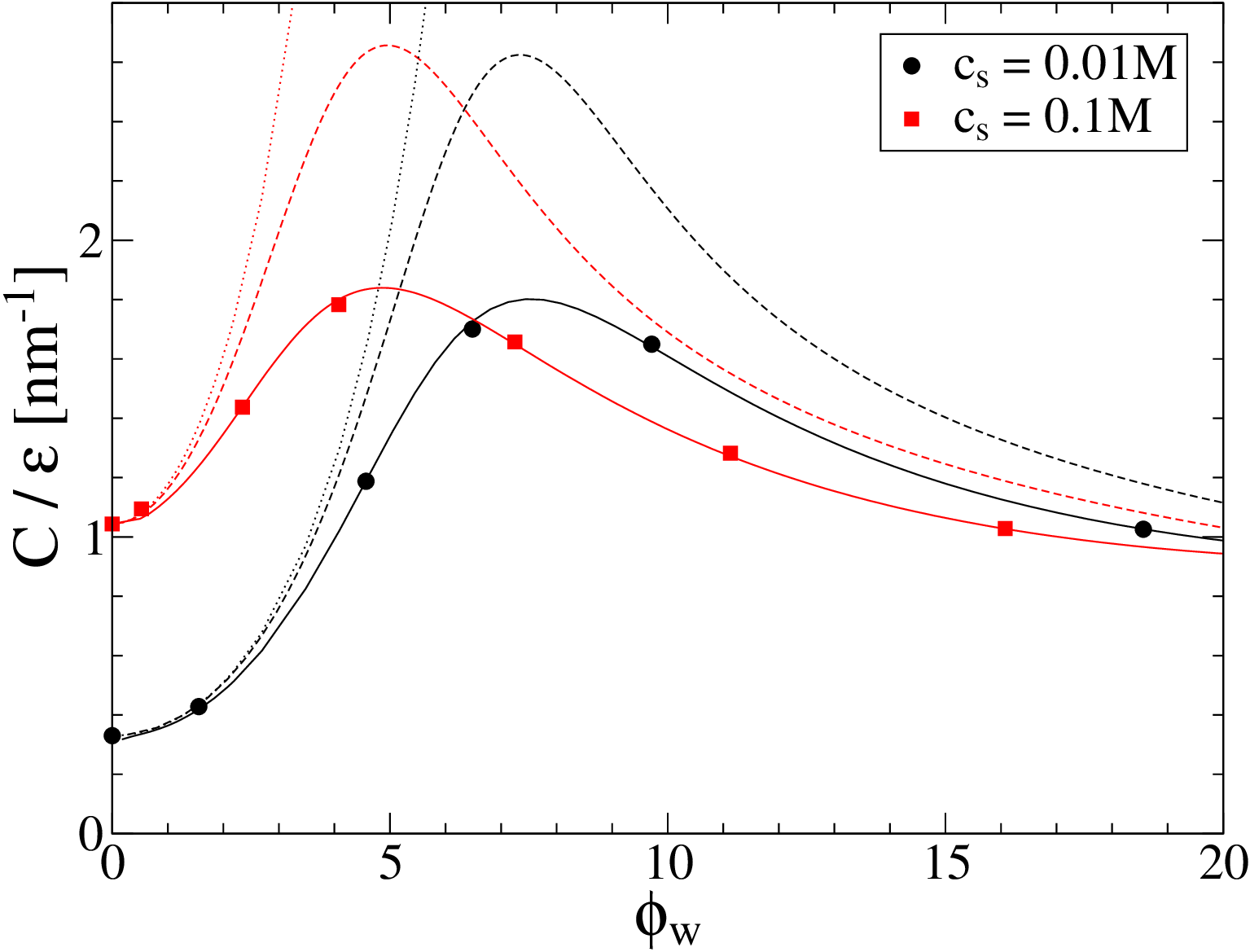}}}
\caption{Capacitance versus the surface potential. The parameters are: $\lambda_B=0.72{\rm nm}$ 
and $\sigma=0.8{\rm nm}$.  The solid lines correspond to the analytical expression, the dashed lines 
are the result of the MPB model, the symbols are the DFT data points, and the dotted lines are the 
PB results, $C_{GC}$.  The agreement between the DFT and the simulation for $C$ was also confirmed in 
\cite{Henderson11}.}
\label{fig:capacitance}
\end{figure}

A quantity that is of particular interest for electrochemistry is the differential capacitance 
of the double-layer \cite{Kornyshev07}, 
$$
C=\frac{\partial\sigma_c}{\partial\psi_w}.
$$
The standard PB result gives
$$
C_{GC} = {\epsilon\kappa}\cosh\Big(\frac{\phi_w}{2}\Big),
$$
which is the, so-called, Gouy-Chapman capacitance.  
$\kappa=\sqrt{8\pi\lambda_Bc_s}$ is the inverse Debye length.  
At low $\phi_w$ this formula works well, 
but as $\phi_w$ increases, steric repulsion lowers the capacitance as counterions are 
prevented from accumulating near the electrode, leading to a 
"camel-shaped" curve of  Fig.~(\ref{fig:capacitance}).  The scaling function derived
in the present paper allows us to easily calculate the differential capacitances, 
which once again are in a very good agreement with the DFT.  

\section{Conclusion}

We have investigated the effects of steric interactions on the ionic distribution and electrostatic 
properties near a charged hard wall.  We have specifically focused on the 
reversible work required to bring a hard spherical particle (with no electric charge) from the bulk 
to the wall, $\Delta_w$.  The quantity $\Delta_w$ is nonlocal and it controls the steric effects in the 
double-layer.  Typically $\Delta_w>0$.  This implies that the overcrowding near a wall causes the depletion 
of counterions away 
from the wall, leading to the increased contact potential.  The less common case, $\Delta_w<0$, represents
the situation where counterions are forced from the dense bulk toward the wall.  This effect decreases 
the value of the contact potential.  

We have also studied the scaling behavior of $\Delta_w$ calculated using the nonlocal DFT
theory.  We expected that the careful look into the scaling behavior can reveal a simple underlying
mechanism.  Unfortunately, no simple model could cover the entire domain.  Instead we proposed a 
functional form to fit the DFT data points for $\Delta_w$.  Such a fit allows for simple and 
accurate calculation of electrostatic properties at contact with the wall without resorting to numerical
schemes of an entire region.  

Finally, we have explored the discrepancies between the LDA and exact description.  The LDA description
shows that $\Delta_w$ depends on two free parameters, in comparison to the three parameters
given by full physical description (DFT).  Another artifact of the LDA is the "saturation" seen in a 
density profile rather than oscillations due to layering.  This saturation effect leads to unphysical 
contact value relation.  For various LDA models, we find that the MPB based on the lattice-gas
model is more accurate than the LDA based on the SPT theory, which for treating a uniform fluid is 
extremely accurate.  This indicates that some cancellation of errors is at work.  

We remind that our analysis is limited to the weak-coupling limit, where electrostatic
interactions are accurately captured by the mean-field treatment.  To accurately account
for the correlations together with the hard-sphere interactions remains a great challenge.  It is our hope
to treat the strong-coupling limit of the hard-sphere ions in the future.

\begin{acknowledgments}
This work was partially supported by the CNPq, Fapergs, INCT-FCx, and by the US-AFOSR 
under the grant FA9550-09-1-0283.  We thank Alexandre Pereira dos Santos for furnishing
us with the simulation data in Fig. (\ref{fig:rho_simulation}).
\end{acknowledgments}




\begin{thebibliography}{99}

\bibitem{Le02}
Y. Levin, Rep. Prog. Phys. {\bf 65},  1577  (2002).

\bibitem{Frydel11}
D. Frydel, J. Chem. Phys. {\bf 134}, 234704 (2011).


\bibitem{Ni59}  E. R. Nightingale, J. Phys. Chem. {\bf 63}, 1381 (1959).

\bibitem{LeSa09} Y. Levin, A. P. dos Santos, A.  Diehl, 
Phys. Rev. Lett.  {\bf 103}, 257802 (2009); A. P. dos Santos,  A. Diehl, 
and Y. Levin,  Langmuir {\bf 26}, 10778 (2010).

\bibitem{Dzubiella10}
I. Kalcher, J. C. F. Schulz, and J. Dzubiella, Phys. Rev. Lett. {\bf 104}, 097802 (2010).

\bibitem{Glueckauf64}
E. Glueckauf, Trans. Faraday Soc. 60, 1637 (1964)

\bibitem{David11}
D. Ben-Yaakov, D. Andelman, and R. Podgornik, J. Chem. Phys. {\bf 134}, 074705 (2011).  

\bibitem{David12}
A. Levy, D. Andelman, H. Orland, http://arxiv.org/abs/1201.6081v1.

\bibitem{Marcia05}
D. Antypov, M. C. Barbosa, and C. Holm, Phys. Rev. E {\bf 71}, 061106 (2005).

\bibitem{Bazant09}
M. Z. Bazant, M S Kilic, B Storey, and A Ajdari, Adv. Colloid Interface Sci., {\bf 152}, 48 (2009).

\bibitem{David97}
I. Borukhov, D. Andelman, H. Orland, Phys. Rev. Lett. {\bf 79}, 435 (1997).

\bibitem{Rosenfeld89}
Y. Rosenfeld, Phys. Rev. Lett. {\bf 63}, 980 (1989).



\bibitem{Lebowitz59}
H. Reiss, H. L. Frisch, J. L. Lebowitz, J. Chem. Phys. {\bf 31}, 369 (1959).

\bibitem{Widom63}
B. Widom, J. Chem. Phys. {\bf 39}, 2808 (1963).

\bibitem{Andrews74}
F. C. Andrews, J. Chem. Phys. {\bf 62}, 272 (1974).

\bibitem{Corti}
M. Heying and D. S. Corti, J. Phys. Chem. B {\bf 108}, 19756 (2004).
D. W. Siderius and D. S. Corti, J. Chem. Phys. {\bf 127}, 144502 (2007).

\bibitem{Roland10}
Roland Roth, J. Phys.: Condens. Matter {\bf 22}, 063102 (2010);
R. Evans, Lecture notes at 3rd Warsaw School of Statistical Physics, Kazimierz Dolny, 
June 2009 pp. 43-85 Warsaw University Press (2010).

\bibitem{Korden12}
S. Korden, http://arxiv.org/abs/1111.5222.  

\bibitem{Rosenfeld96}
Y. Rosenfeld, M. Schmidtz, H. L\"owen, and P. Tarazona, 
J. Phys.: Condens. Matter {\bf 8}, L577 (1996).


\bibitem{Yan12}
A. Diehl, A. P. dos Santos, Y. Levin, J. Phys.: Condens. Matter {\bf 24}, 284115 (2012).

\bibitem{Kornyshev07}
A. A. Kornyshev, J. Phys. Chem. B {\bf 111}, 5545 (2007). 


\bibitem{Frydel99}
D. Frydel, M. Oettel, S. Dietrich, Phys. Rev. Lett. {\bf 99}, 118302 (2007);
D. Frydel and M. Oettel, Phys. Chem. Chem. Phys. {\bf 13}, 4109 (2011).



\bibitem{Lebowitz79}
D. Henderson, L. Blum, J. L. Lebowitz, J. Electroanal. Chem. {\bf 102}, 315 (1979).


\bibitem{SaDi09} 
A. P. dos Santos, A. Diehl, and Y. Levin, J. Chem. Phys. {\bf 130}, 124110 (2009).

\bibitem{Henderson11}
D. Henderson, S. Lamperski, Z. Jin, and J. Wu, J. Phys. Chem B {\bf 115}, 12911 (2011).






\end{thebibliography}
\end{document}